 \documentstyle[prb,aps,floats,epsf,twoside]{revtex}

\begin{document}

  \twocolumn[\hsize\textwidth\columnwidth\hsize\csname
  @twocolumnfalse\endcsname

\draft
\preprint{Submitted to Phys. Rev. B}

\title{Conduction band spin  splitting and negative  magnetoresistance
in ${\rm A}_3{\rm B}_5$ heterostructures.}

\author{F. G. Pikus}

\address{Department  of Physics,  University of  California at  Santa
Barbara, Santa Barbara, CA 93106}

\author{G. E. Pikus}

\address{A.  F.  Ioffe  Physico-Technical  Institute,  194021,   St.
Petersburg, Russia.}

\date{\today}

\maketitle

\begin{abstract}

The  quantum   interference  corrections   to  the   conductivity  are
calculated  for  an  electron  gas  in  asymmetric  quantum wells in a
magnetic field. The theory takes  into account two different types  of
the spin splitting of the conduction band: the Dresselhaus terms, both
linear and cubic in  the wave vector, and  the Rashba term, linear  in
wave vector. It  is shown that  the contributions of  these terms into
magnetoconductivity are not additive, as it was traditionally assumed.
While the contributions of all terms of the conduction band  splitting
into the D'yakonov--Perel' spin  relaxation rate are additive,  in the
conductivity the two  linear terms cancel  each other, and,  when they
are equal, in the absence of the cubic terms the conduction band  spin
splitting does  not show  up in  the magnetoconductivity  at all.  The
theory agrees very well with  experimental results and enables one  to
determine experimentally parameters of the spin-orbit splitting of the
conduction band.

\end{abstract}


  \vskip 2pc ] 

\narrowtext

It was first found by Dresselhaus \cite{Dresselhaus55} in 1955 that in
cubic crystals with  symmetry $T_d$ there  is a spin  splitting of the
conduction band, which is cubic in the electron wave vector $k$.  This
splitting is described by the Hamiltonian\cite{Rashba61,Dyakonov71}:

\begin{eqnarray}
H_s = & & \gamma \sum_i \sigma_i k_i \left(k_{i+1}^2-k_{i+2}^2\right),
\nonumber \\
& & \left( i = x, y, z;\quad i + 3 \rightarrow i \right),
\label{HamBulk}
\end{eqnarray}

\noindent where $\sigma_i$  are the Pauli  matrices (in this  paper we
take $\hbar = 1$ everywhere except in final formulas). If the symmetry
of a crystal is reduced, the splitting linear in wave vector  appears,
for example, in uniaxial  crystals \cite{Rashba59}, in $T_d$  crystals
under a deformation\cite{Bir61,PikusBook},  and, most importantly,  in
quantum             wells             and             heterostructures
\cite{PikusBook,Altshuler81,Bychkov84,Dyakonov86,Lyanda94}.         In
symmetric  quantum  wells  grown  along  $[001]$,  the conduction band
Hamiltonian has the form:

\begin{equation}
H = {k^2 \over 2 m^*} + \left(\mbox{\boldmath $\sigma$}{\bf \Omega}
\right),
\label{HamSymm}
\end{equation}

\noindent where  $\mbox{\boldmath $\sigma$}  = (\sigma_x,  \sigma_y)$,
${\bf \Omega} = (\Omega_x, \Omega_y)$ are two-dimensional vectors with
components in the plane of  the quantum well. Here the  spin splitting
coefficients ${\bf \Omega}$ are  proportional to the bulk  coefficient
$\gamma$       in       Eq.~(\ref{HamBulk}).       According        to
Ref.~\onlinecite{Iordanskii94},

\begin{eqnarray}
\Omega_x &=& - \Omega_1^{(1)} \cos \varphi - \Omega_3 \cos 3\varphi,
\nonumber \\
\Omega_y &=& \ \Omega_1^{(1)} \sin \varphi - \Omega_3 \sin 3\varphi,
\label{Omega} \\
\Omega_1^{(1)} &=& \gamma k \left( \left\langle k_z^2 \right\rangle -
{1 \over 4} k^2 \right),
\quad
\Omega_3 = \gamma {k^3 \over 4},
\nonumber
\end{eqnarray}

\noindent where $k^2 = k_x^2  + k_y^2$, $\tan \varphi =  k_x/k_y$, and
$\left\langle k_z^2 \right\rangle$ is  the average wave vector  in the
direction $z$, normal to the quantum well.

In  asymmetric  quantum  wells,  or  in  the presence of a deformation
$\epsilon_{xy}$,  the  Hamiltonian  $H$  includes  another,  so-called
Rashba term\cite{Bychkov84}:

\begin{equation}
H^\prime = \alpha [\mbox{\boldmath $\sigma k$}]_z.
\label{HamAsymm}
\end{equation}

\noindent   This   term   can   be   included   in   the   Hamiltonian
Eq.~(\ref{HamSymm})  if  one  includes  additional  terms  into  ${\bf
\Omega}$:

\begin{equation}
\Omega_x = \Omega_1^{(2)} \sin \varphi, \quad
\Omega_y = -\Omega_1^{(2)} \cos \varphi, \quad
\Omega_1^{(2)} = \alpha k.
\label{Omega2}
\end{equation}

\noindent      In      deformed      crystals,      according       to
Refs.~\onlinecite{Bir61,PikusBook,Pikus84},

\begin{equation}
\alpha = {1 \over 2} C_3 \epsilon_{xy}.
\label{deform}
\end{equation}

\noindent The values of the coefficient $C_3$ for some ${\rm A}_3 {\rm
B}_5$ compounds are given in Refs.~\onlinecite{PikusBook,Pikus84}.  In
an asymmetric quantum well the coefficient $\alpha$ is proportional to
the average value of the electric field (or potential gradient) in the
well.  This  coefficient  was  calculated  in  the  framework  of  the
$\mbox{\boldmath      $k$}\cdot\mbox{\boldmath      $p$}$       method
\cite{Lommer88,Andrada94}, neglecting  the immediate  vicinity of  the
potential barriers. However, if the effective mass approximation would
have been valid  throughout the entire  well, including the  barriers,
then $\alpha = 0$.

If the linear in $k$ spin splitting is given by only one of the  terms
Eq.~(\ref{Omega}) or  Eq.~(\ref{Omega2}), all  observable effects  are
identical, because these  two Hamiltonians can  be converted one  into
the other by  a unitary transformation.  In both cases  the conduction
band minimum  is shaped  like a  ring around  $k=0$. However,  if both
terms are  present, the  electron spectrum  changes qualitatively: the
energy  minima  now  occur  at  finite  $k$  along the axes $(110)$ or
$(1\bar{1}0)$,  depending  on   the  signs  of   $\Omega_1^{(1)}$  and
$\Omega_1^{(2)}$.

Both  terms  Eq.~(\ref{Omega})  and  Eq.~(\ref{Omega2})  give additive
contributions    into     the    D'yakonov--Perel'     spin-relaxation
rate\cite{Dyakonov86}:

\begin{equation}
{1 \over \tau_{s_{xx}}} =
{1 \over 2 \tau_{s_{zz}}} =
2 \left(\Omega_1^2 \tau_1 + \Omega_3^2 \tau_3 \right),
\label{spin}
\end{equation}

\noindent where $\Omega_1^2 = {\Omega_1^{(1)}}^2 + {\Omega_1^{(2)}}^2$
and $\tau_n$, $n  = 1, 3$,  is the relaxation  time of the  respective
component of the distribution function:

\begin{equation}
{1 \over \tau_n} = \int W(\vartheta) \left( 1 - \cos n\vartheta
\right) \,d\vartheta.
\end{equation}

\noindent Here $W(\vartheta)$ is  the probability of scattering  by an
angle $\vartheta$.

In this paper  we show that  such additivity does  not exist for  weak
localization  effects,   which  are   responsible  for   the  negative
magnetoresistance (NMR). In the theory  of the NMR the spin  splitting
of   the   conduction   band   was   first   taken   into  account  in
Ref.~\onlinecite{Altshuler81}. In this paper it was supposed that  the
magnetoconductivity  $\Delta\sigma(B)$  depends   only  on  the   spin
relaxation   times,   by   analogy   with   the  Larkin-Hikami-Nagaoka
theory\cite{Hikami80},    which    considered    the     Elliott-Yafet
spin-relaxation  mechanism.  In  Ref.~\onlinecite{Iordanskii94} it was
shown  that  for  D'yakonov--Perel'  spin  relaxation this approach is
valid when Hamiltonian Eq.~(\ref{HamSymm}) contains only cubic in  $k$
terms, the ones  with $\Omega_3$ (note  also that the  spin relaxation
rates, given in Ref.~\onlinecite{Altshuler81}, should be increased two
times\cite{Iordanskii94}).      The      formulas      derived      in
Ref.~\onlinecite{Iordanskii94} can be  used if only  one of the  terms
Eq.~(\ref{Omega}) or Eq.~(\ref{Omega2}) is present in the  Hamiltonian
Eq.~(\ref{HamSymm}).

When both terms Eq.~(\ref{Omega}) or Eq.~(\ref{Omega2}) coexist in the
conduction  band  Hamiltonian,  one  can  reduce  the equation for the
Cooperon propagator  ${{\rm \kern.24em  \vrule width.05em  height1.4ex
depth-.05ex \kern-.26em C}}_0(\mbox{\boldmath $q$})$, as described  in
Ref.~\onlinecite{Iordanskii94}, using the iteration in the  parameters
$\Omega_i  \tau_0$   and  $\mbox{\boldmath   $q$}\mbox{\boldmath  $v$}
\tau_0$, where $\tau_0^{-1} = \int W(\vartheta) \,d \vartheta$ is  the
elastic lifetime and $v$ is the Fermi velocity, to the following form:

\begin{equation}
{\cal H}
{{\rm \kern.24em
            \vrule width.05em height1.4ex depth-.05ex
            \kern-.26em C}}_0
= {1 \over 2 \pi \nu_0 \tau_0^2},
\label{Cooperon}
\end{equation}

\noindent where $\nu_0$  is the density  of states at  the Fermi level
and

\begin{eqnarray}
{\cal H} = & &{1 \over \tau_\varphi} + {1 \over 2} v^2 q^2 \tau_1 +
\nonumber \\
& & \left(\Omega_1^2\tau_1 + \Omega_3^2\tau_3\right)
\left(2+\sigma_x \rho_x + \sigma_y \rho_y \right) +
\nonumber \\
& & 2 \left(\sigma_x \rho_y + \sigma_y \rho_x\right) \Omega_1^{(1)}
\Omega_1^{(2)}\tau_1 +
\label{CoopH} \\
& &
v \tau_1 \left[ \left(\sigma_x + \rho_x\right)
\left(-\Omega_1^{(1)} q_x + \Omega_1^{(2)} q_y \right) +
\right.
\nonumber \\
& & \left.
\left(\sigma_y + \rho_y\right)
\left(\Omega_1^{(1)} q_y - \Omega_1^{(2)} q_x \right) \right].
\nonumber
\end{eqnarray}

\noindent  The  components  of  the  matrix  ${{\rm  \kern.24em \vrule
width.05em height1.4ex depth-.05ex  \kern-.26em C}}_0$ are  determined
by two  pairs of  spin indices,  and the  matrices \boldmath  $\sigma$
\unboldmath  act  on  the  first  pair,  while  the matrices \boldmath
$\rho$\unboldmath,  also  Pauli  matrices,  -  on the second pair. The
magnetic fiend dependent correction to the conductivity is  determined
by the quantity $S(\mbox{\boldmath $q$})$ \cite{Iordanskii94}:

\begin{equation}
\Delta \sigma = - {e^2 D  \over 4 \pi^3}
\int\limits_0^{q_{max}} S(\mbox{\boldmath $q$}) \,d^2 q
\label{NMR}
\end{equation}

\noindent where $q_{max}^2 = (D  \tau_1)^{-1}$, $D = v^2 \tau_1/2$  is
the diffusion coefficient, and

\begin{equation}
S(\mbox{\boldmath $q$}) = 2 \pi \nu_0 \tau_0^2 \sum_{\alpha \beta}
{{\rm  \kern.24em \vrule width.05em
height1.4ex depth-.05ex  \kern-.26em C}}
_{0_{\alpha \beta \beta \alpha}} =
- {1 \over {\cal E}_0} + \sum^1_{m = -1} {1 \over E_m},
\label{S}
\end{equation}

\noindent ${\cal E}_0$ and $E_m$  are the eigenvalues of the  operator
${\cal   H}$,   corresponding    to   the   eigenfunctions    $\phi_0$
(antisymmetric singlet state) and $\phi_l^m$ with $l = 1$, $m = -1, 0,
1$ (symmetric triplet state). The singlet contribution is the same  as
in Ref.~\onlinecite{Iordanskii94}:

\begin{equation}
{\cal E}_0 = D q^2 + {1 \over \tau_\varphi}.
\label{E0}
\end{equation}

The values $E_m$ are the eigenvalues of the matrix operator

\begin{eqnarray}
\tilde{\cal H} = & &D q^2 + {1 \over \tau_\varphi} +
2 \left(\Omega_1^2\tau_1 + \Omega_3^2\tau_3\right)
\left(2 - J_z^2 \right) -
\nonumber \\
& & 4 i \Omega_1^{(1)} \Omega_1^{(2)}\tau_1
\left( J_+^2 - J_-^2 \right) +
\label{Er} \\
& &
2 (D \tau_1)^{1/2}
\left[-\Omega_1^{(1)} \left(J_+ q_+ + J_- q_- \right) +
\right.
\nonumber \\
& & \left.
      i\Omega_1^{(2)} \left(J_+ q_- - J_- q_+ \right) \right].
\nonumber
\end{eqnarray}

\noindent Here $J_i$ are the matrices of the angular momentum operator
with  total  momentum  $L=1$,  $J_\pm  = (J_x \pm iJ_y)/\sqrt{2}$, and
$q_\pm = q_x \pm i q_y$.

An  interesting  particular  case  occurs  when  $\Omega_1^{(1)} = \pm
\Omega_1^{(2)}$ and $\Omega_3 = 0$. If one introduces new  coordinates
(upper sign for $\Omega_1^{(1)} = + \Omega_1^{(2)}$ and lower sign for
$\Omega_1^{(1)} = - \Omega_1^{(2)}$)

\begin{eqnarray}
& & v = {x \pm y \over \sqrt{2}}, \
u = {x \mp y \over \sqrt{2}}, \nonumber \\
& & q_v = {q_x \pm q_y \over \sqrt{2}}, \
q_u = {q_x \mp q_y \over \sqrt{2}}, \label{uv} \\
& & J_v = {J_x \pm J_y \over \sqrt{2}}, \
J_u = {J_x \mp J_y \over \sqrt{2}}, \nonumber
\end{eqnarray}

\noindent it  is easy  to show  that the  operator $\tilde{\cal H}$ in
these coordinates becomes

\begin{eqnarray}
\tilde{\cal H} & = & D \left[ q_v^2  +
\left(q_u - q_{u0} J_v \right)^2\right] + {1 \over \tau_\varphi}, \\
q_{u0}^2 & = & {4 \Omega_1^2 \tau_1 \over D} \nonumber
\end{eqnarray}

\noindent  In  the  basis  of  eigenfunctions  of  $J_v$ we have three
independent equations for eigenvalues $E_m$:

\begin{equation}
E_m = D \left[ q_v^2  +
\left(q_u + m q_{u0} \right)^2\right] + {1 \over \tau_\varphi}.
\label{ErSplit}
\end{equation}

\noindent In this case $S(\mbox{\boldmath $q$})$ becomes

\begin{eqnarray}
S(\mbox{\boldmath $q$})  =  & &
\left\{D \left[ q_v^2 + \left(q_u + q_{u0} \right)^2\right] + {1 \over
\tau_\varphi}\right\}^{-1} + \nonumber \\
& & \left\{D \left[ q_v^2 + \left(q_u - q_{u0} \right)^2\right] + {1
\over \tau_\varphi}\right\}^{-1}.
\end{eqnarray}

\noindent When calculating  the conductivity Eq.~(\ref{NMR}),  one can
neglect the shift in $q$-space  $q_{u0}$ compared to $q_{max}$ on  the
upper  limit  of  the  integral.  The  result  for  the   conductivity
correction is:

\begin{equation}
\Delta \sigma =
-{e^2 \over 2 \pi^2 \hbar } \ln {\tau_\varphi \over \tau_1}.
\end{equation}

\noindent Note that in order for the diffusion approximation itself to
be valid, the condition $\tau_\varphi  \gg \tau_1$ must hold. One  can
see that when  $\Omega_1^{(1)} = \pm  \Omega_1^{(2)}$ and $\Omega_3  =
0$, $\Delta \sigma$ is determined  by the same formula as  without any
spin relaxation. Note that  the spin relaxation rate  Eq.~(\ref{spin})
does not show any peculiar behavior in this case.

The  reason  for  such  a  striking  difference  between  NMR and spin
relaxation  can  be  seen  if  one  writes the Hamiltonian $H^\prime =
\mbox{\boldmath  $\sigma$}  {\bf  \Omega}$  at  $\Omega_1^{(1)}  = \pm
\Omega_1^{(2)}$ and $\Omega_3 = 0$ in the form

\begin{equation}
H^\prime = 2 \alpha (\sigma_v k_u),
\label{SimpleHam}
\end{equation}

\noindent where  the coordinates  $u$ and  $v$ are  determined by  Eq.
(\ref{uv}).  Then  one  can  see  that the random ``effective magnetic
field'' $2  \alpha k_u$,  parallel to  axis $v$,  leads to  the random
precession of electron spin in the plane, perpendicular to this  axis.
The frequency  of this  rotation is  proportional to  $k_u$, or to the
velocity $V_u$.  During the  time $\delta  t$ between  two consecutive
scattering  acts  the  spin   rotates  by  an  angle   $\varphi_{21}$,
proportional to the  distance $V_u \delta  t = u_2  - u_1$. Therefore,
the total  angle of  the spin  precession along  a path between points
with coordinates $u_0$ and $u_n$ is proportional to the length $L_n  =
u_n - u_0$. For the spin  relaxation, these paths can be anything  and
resulting spin  rotations are  random. On  the other  hand, the NMR is
determined only by  the trajectories for  which $L_n$ is  smaller than
the electron wavelength,  i.e. $L_n =  0$ within the  framework of our
theory. Hence, for $\Omega_1^{(1)} = \pm \Omega_1^{(2)}$ and $\Omega_3
= 0$ the total spin rotation {\em on the trajectories which contribute
into the NMR}\/ is zero\cite{DyakonovHelp} When $\Omega_1^{(1)} =  \pm
\Omega_1^{(2)}$, but  $\Omega_3 \ne  0$, the  direction of  the random
magnetic field changes with the direction of \mbox{\boldmath $k$}, and
the total spin rotation on a  closed trajectory is not zero any  more.
The  absence  of  a  spin  contribution  to  interference  effects  in
one-dimensional  metal  rings  for  a  spin  Hamiltonian,  similar  to
Eq.~(\ref{SimpleHam}),      has      been      pointed      out     in
Refs.~\onlinecite{Lyanda94,LyandaPress},   where   the    conductivity
oscillations  and  universal  fluctuations  of  the  conductance  were
considered.

In a  magnetic field  $q_\pm$ become  operators and  can be  expressed
through the operators $a^\dagger$ and $a$, which increase and decrease
the Landau level number $n$\cite{Iordanskii94}:

\begin{eqnarray}
D^{1/2} q_+ = \delta^{1/2} a,
\quad
D^{1/2} q_- & = & \delta^{1/2} a^\dagger,
\quad
D q^2 = \delta \{a a^\dagger\},
\nonumber \\
\{a a^\dagger\} & = & {1 \over 2} (a a^\dagger + a^\dagger a),
\label{me}
\end{eqnarray}

\noindent where

\begin{equation}
\delta = {4 e B D \over \hbar c}.
\label{delta}
\end{equation}

\noindent The non-zero matrix elements of these operators are

\begin{eqnarray}
\left\langle n-1 \right| a \left| n \right\rangle & = &
\left\langle n \right| a^\dagger \left| n-1 \right\rangle = \sqrt{n},
\nonumber \\
\left\langle n \right| \{a a^\dagger\} \left| n \right\rangle
& = & n + {1 \over 2}.
\end{eqnarray}

\noindent The magnetic field dependent correction to the  conductivity
Eq.~(\ref{NMR}) now becomes

\begin{equation}
\Delta \sigma = - {e^2 \delta \over 4 \pi^2 \hbar} \sum_{n=0}^{n_{max}}
\left(- {1 \over {\cal E}_{0n}} + \sum_{m=-1}^1 {1 \over E_{mn}}\right),
\label{NMRB}
\end{equation}

\noindent   where   $n_{max}    =   1/\delta\tau_1$.   According    to
Eqs.~(\ref{E0},\ref{Er},\ref{me}),

\begin{equation}
{\cal E}_{0n} = \delta \left(n + {1 \over 2} \right) +
{1 \over \tau_\varphi},
\label{E0Mag}
\end{equation}

\noindent and $E_{mn}$ are the eigenvalues of the operator

\begin{eqnarray}
\tilde{\cal H} = & & \delta \{a a^\dagger\} + {1 \over \tau_\varphi} +
2 \left(\Omega_1^2\tau_1 + \Omega_3^2\tau_3\right)
\left(2 - J_z^2 \right) -
\nonumber \\
& & 4 i \Omega_1^{(1)} \Omega_1^{(2)}\tau_1
\left( J_+^2 - J_-^2 \right) + 2 (\delta \tau_1)^{-1}
\label{ErMag} \\
& &
\left[-\Omega_1^{(1)} \left(J_+ a - J_- a^\dagger \right) +
      i\Omega_1^{(2)} \left(J_+ a^\dagger - J_- a \right) \right].
\nonumber
\end{eqnarray}

One can see that in  the general case, when both  $\Omega_1^{(1)}$ and
$\Omega_1^{(2)}$ are non-zero, the determinant of $\tilde{\cal H}$ can
no longer  be split  into submatrices  $3 \times  3$ in  the basis  of
functions $\left|n, m\right\rangle$ ($m = -1, 0, 1$), unlike  Eq.~(10)
of  Ref.~\onlinecite{Iordanskii94}.  Therefore,  the  only way to find
eigenvalues  $E_{mn}$  is   to  diagonalize  numerically   the  matrix
$\tilde{\cal H}$.\footnote{Since we are interested only in the sum  of
reciprocal eigenvalues, it is possible to express it through minors of
the  matrix  $\tilde{\cal   H}$  without  computing   the  eigenvalues
themselves. However, this procedure  has about the same  complexity as
full diagonalization.} The  number of elements  one has to  take for a
given value of magnetic field $B$, or $\delta$, is at least  $n_{max}$
and increases infinitely as $B$ approaches $0$. Note that the size  of
the matrix $\tilde{\cal H}$ is $N = 3 n_{max}$.

The  numerical  diagonalization  of  the  matrix  $\tilde{\cal H}$ was
performed using the LAPACK eigensolver for hermitian banded  matrices.
To improve  convergence of  the sum  in Eq.~(\ref{NMRB}),  we add  and
subtract  from  each  $1/E_{mn}$  its  approximate  asymptotic   value
$1/\delta(n + 1)$  at large $n$  (the constant, $1$,  added to $n$  is
needed only to extend summation to $n=0$, as in Eq.~(\ref{NMRB})). The
sum of  terms $\delta  E_{mn}^{-1}-(n+1)^{-1}$ can  be extended  to $n
\rightarrow \infty$, while the sum of added terms $(n+1)^{-1}$ can  be
replaced by an integral. Both approximations cause errors proportional
to $\tau_1/\tau_\varphi$, $(\Omega_i \tau_i)^2$, and $1/n_{max}$,  but
the very approach of this paper  is valid only when this parameter  is
small and  $n_{max}$ is  very large.  Therefore, we  use the following
expression for the magnetoconductivity:

\begin{eqnarray}
\Delta \sigma = & & - {e^2 \over 4 \pi^2 \hbar}
\left\{
\sum_{n=0}^{\infty} \left[- {\delta \over {\cal E}_{0n}} + {1 \over n +
1} \right] +
\right.  \nonumber \\
& & \sum_{n=0}^{\infty} \sum_{m=-1}^1 \left[ {\delta \over E_{mn}} -
{1 \over n + 1} \right]   -
\label{NMRBConv} \\
& & 2 \ln (\delta \tau_1) - 2C \Biggr\},
\nonumber
\end{eqnarray}

\noindent where $C$ is the Euler constant. In order to compute the sum
of $1/E_{mn}$ to $n = \infty$ numerically, we perform the calculations
for  few  values  of  $n_{max}$  in  the  range  from  500 to 5000 and
extrapolate to $n \rightarrow \infty$.

\begin{figure}[t]

\epsfxsize=3 in
\epsffile{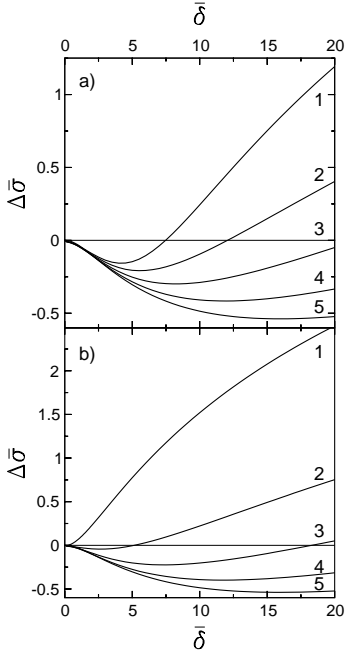}

\caption{Dimensionless  magnetoconductivity  $\Delta\bar\sigma$   {\it
vs}\/       dimensionless       magnetic       field      $\bar\delta$
{\bf(}Eq.~(\protect\ref{dimen}){\bf)} for  $H_{SO}/H_\varphi =  4$ and
$\Omega_1^{(2)} =  0$ (a)  and $\Omega_1^{(1)}  = \Omega_1^{(2)}$ (b).
For both plots (a) and (b) the curves 1 through 5 show dependencies at
different  $H^\prime_{SO}/H_{SO}   =  1$,   3/4,  1/2,   1/4,  and  0,
respectively. \label{Compare}}

\end{figure}

We  now  present  the  results  of  the numerical computations. Let us
introduce       the       following       characteristic      magnetic
fields\cite{Iordanskii94}:

\begin{eqnarray}
H_\varphi & = & {c \hbar \over 4 e D \tau_\varphi}, \
{B \over H_\varphi} = \delta\tau_\varphi,
\nonumber \\
H_{SO} & = & {c \hbar \over 4 e D} \left(2\Omega_1^2\tau_1 +
2\Omega_3^2 \tau_3 \right), \
H^\prime_{SO}   =   {c \hbar \over 4 e D} 2\Omega_1^2\tau_1,
\label{HSO} \\
H^{(1)}_{SO} & = & {c \hbar \over 4 e D} {2\Omega_1^{(1)}}^2\tau_1, \
H^{(2)}_{SO}   =   {c \hbar \over 4 e D} {2\Omega_1^{(2)}}^2\tau_1.
\nonumber
\end{eqnarray}

\noindent Note  that the  field $H_{SO}$  is proportional  to the spin
relaxation rate. We also use dimensionless units for the  conductivity
and magnetic field:

\begin{eqnarray}
\bar\delta & \equiv & {B \over H_\varphi} = \delta\tau_\varphi,
\nonumber \\
\Delta\bar\sigma & = & {4 \pi^2 \hbar \over e^2}
\left(\Delta\sigma - \lim_{\delta \rightarrow 0}\Delta\sigma\right).
\label{dimen}
\end{eqnarray}

\begin{figure}[t]

\epsfxsize=3 in
\epsffile{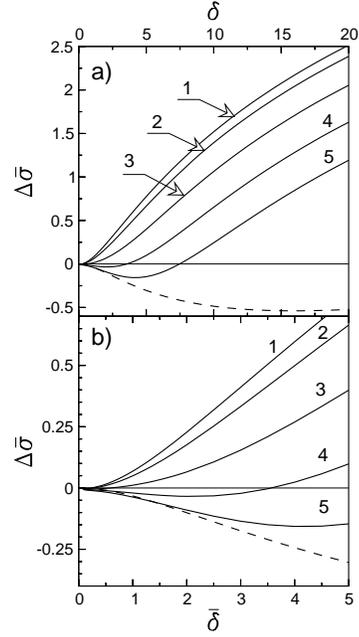}

\caption{Dimensionless  magnetoconductivity  $\Delta\bar\sigma$   {\it
vs}\/       dimensionless       magnetic       field      $\bar\delta$
{\bf(}Eq.~(\protect\ref{dimen}){\bf)} for  $H_{SO}/H_\varphi =  4$ and
for $H^\prime_{SO}/H_{SO} = 1$ ($\Omega_3=0$). The curves 1 through  5
show dependencies  at different  ratios $H^{(1)}_{SO}/H^\prime_{SO}  =
1/2$,  5/8,  3/4,   7/8,  and  1,   respectively.  Lower  plot   shows
magnification of small magnetic fields region. The dashed curve  shows
the dependency for $\Omega_1=0$. \label{Redistr}}

\end{figure}

We begin by demonstrating the effect of the coexistence of both  terms
$\Omega_1^{(1)}$  and  $\Omega_1^{(2)}$  in  the  spin  splitting.  In
Fig.~\ref{Compare}a      we      reproduce      the     results     of
Ref.~\onlinecite{Iordanskii94}     for     magnetoconductivity      at
$H_{SO}/H_\varphi  =  4$  and  different $H^\prime_{SO}/H_{SO}$. These
results  can  be  obtained  from  Eqs.~(\ref{NMRB}-\ref{ErMag}) if one
leaves only $\Omega_1^{(1)}$, or $\Omega_1^{(2)}$, and sets the  other
one to 0. Our results  have better numerical accuracy, especially  for
small $\bar\delta$, due to extrapolation to $n \rightarrow \infty$  in
the  sum  Eq.~(\ref{NMRBConv}).  Note  that  the  lowest  curve,  with
$H^\prime_{SO}=0$,  gives  the  result  of  the  Larkin-Hikami-Nagaoka
theory\cite{Hikami80}. In Fig.~\ref{Compare}b  we show the  curves for
the same values of $H_{SO}/H_\varphi$ and $H^\prime_{SO}/H_{SO}$,  but
now $\Omega_1^{(1)} = \Omega_1^{(2)}$. The effect of redistribution of
the spin splitting  between $\Omega_1^{(1)}$ and  $\Omega_1^{(2)}$ is,
naturally, more pronounced for large $H^\prime_{SO}/H_{SO}$, when  the
linear in $k$ term dominates the spin relaxation. One can see that for
$H^\prime_{SO}/H_{SO} > 0.5$ the results in Fig.~\ref{Compare} a and b
are  qualitatively  different:  the  magnetoresistance  minimum shifts
closer to $B=0$ and eventually disappears, and $\Delta\sigma$  becomes
monotonic.

This effect is shown in more details in Fig.~\ref{Redistr}, where  the
magnetoconductivity  is  plotted  for  $H^\prime_{SO}/H_{SO}  = 1$ and
various   $H^{(1)}_{SO}/H^\prime_{SO}$.   The   lowest   solid   curve
reproduces  the  result  of  Ref.~\onlinecite{Iordanskii94}.  One  can
clearly  see  the  shift   of  magnetoconductivity  minimum  and   its
disappearance  when   $\Omega_1^{(1)}$  and   $\Omega_1^{(2)}$  become
comparable ($H^{(1)}_{SO}/H^\prime_{SO}$ close to $1/2$). The  minimum
disappears and the slope of magnetoconductivity at $B=0$ changes  sign
at $H^{(1)}_{SO}/H^\prime_{SO} \approx 3/4$.

\begin{figure}[t]

\epsfxsize=3 in
\epsfysize=3 in
\epsffile{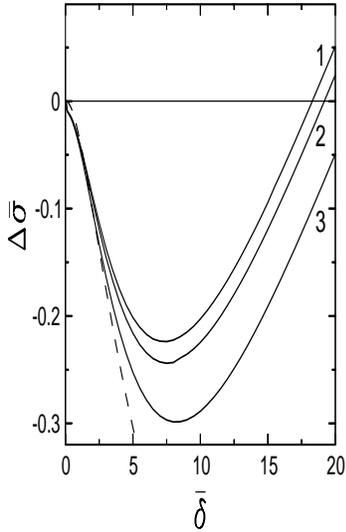}

\caption{Dimensionless  magnetoconductivity  $\Delta\bar\sigma$   {\it
vs}\/       dimensionless       magnetic       field      $\bar\delta$
{\bf(}Eq.~(\protect\ref{dimen}){\bf)} for  $H_{SO}/H_\varphi =  4$ and
for $H^\prime_{SO}/H_{SO}  = 1/2$.  The curves  1, 2,  3 correspond to
$H^{(1)}_{SO}/H^\prime_{SO}  =  1/2$,  3/4,  and  1, respectively. The
dashed curve shows the dependency for $\Omega_1=0$. \label{Smaller}}

\end{figure}

While  redistribution  of  linear  in  $k$  spin splitting between the
Dresselhaus    term    Eq.~(\ref{Omega})    and    the   Rashba   term
Eq.~(\ref{Omega2}) has maximum effect on the magnetoconductivity  when
the linear  splitting is  dominant, the  quantitative effects  of such
redistribution  can  be  seen  when  linear  and  cubic splittings are
comparable.     In     Fig.~\ref{Smaller}     the     dependence    of
magnetoconductivity  on  $H^{(1)}_{SO}/H^\prime_{SO}$  is  shown   for
$H^\prime_{SO}/H_{SO}=1/2$, when the contributions of both linear  and
cubic terms to the spin relaxation  rate are equal. One can see  that,
while the effect is not  as dramatic as in Fig.~\ref{Redistr},  it has
qualitatively the same character.

We now return to  the question of the  cancellation of the Rashba  and
Dresselhaus  terms  in  linear  spin  splitting.  One can see that the
cancellation of  spin relaxation  terms in  conductivity, which occurs
when  $\Omega_1^{(1)}  =  \Omega_1^{(2)}$  and  $\Omega_3  =  0$, also
happens in  a magnetic  field. In  this case  the eigenvalue  equation
Eq.~(\ref{ErMag}) splits into  three independent equations,  analogous
to Eq.~(\ref{ErSplit}).  The commutation  relations for  the operators
$q_u$ and $q_v$ in  a magnetic field do  not change with the  shift of
$q_u$ by a constant value $q_{u0}$ in each of these equations:

\begin{equation}
D \left[ q_v, q_u \pm q_{u0} \right]
 = -i\, {\delta \over 2}.
\end{equation}

Therefore, all eigenvalues $E_{mn}$ are equal to ${\cal E}_{0n}$, and

\begin{equation}
\Delta \bar{\sigma} =
- 2 \left[ \psi\left({1 \over 2} + {H_\varphi \over B}\right) -
\ln{H_\varphi \over B}\right],
\end{equation}

\noindent where $\psi$ is a digamma-function.

\begin{figure}[t]

\epsfxsize=3 in
\epsffile{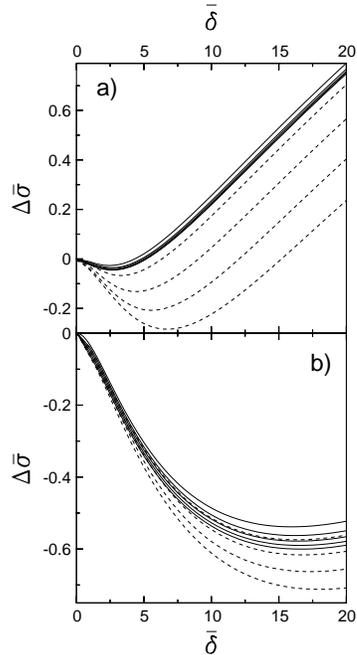}

\caption{Cancellation of linear terms in spin splitting. Dimensionless
magnetoconductivity   $\Delta\bar\sigma$   is   plotted   {\it   vs}\/
dimensionless           magnetic           field          $\bar\delta$
{\bf(}Eq.~(\protect\ref{dimen}){\bf)} for constant $\Omega_3^2  \tau_3
\tau_\varphi = 1/2$ for plot a) and $2$ for plot b), and for different
$H^\prime_{SO}/H_\varphi$. Solid lines show the magnetoresistance when
$\Omega_1^{(1)} = \Omega_1^{(2)}$ (maximum cancellation), dashed lines
are for  $\Omega_1^{(2)} =  0$ (no  cancellation). For  each family of
curves (solid or dashed) $H^\prime_{SO}/H_\varphi$ takes values 0,  1,
2, 3, and 4, with 0 for the uppermost curve and 4 for the lowest curve  (at
$H^\prime_{SO}$ solid and dashed curves coincide). \label{Cancel}}

\end{figure}

As  we   have  already   noted,  this   cancellation  of   terms  with
$\Omega_1^{(1)}$ and $\Omega_1^{(2)}$  occurs only when  $\Omega_3=0$.
However, it is  reasonable to suppose  that addition of  a small cubic
splitting will break this  cancellation only slightly, resulting  in a
very weak  dependence of  the magnetoconductivity  on $\Omega_1$  when
$\Omega_1^{(1)} = \Omega_1^{(2)}$. Fig.~\ref{Cancel} shows that it  is
indeed so. In Fig.~\ref{Cancel}a the magnetoconductivity is  presented
for small $\Omega_3$ and different  $\Omega_1$. One can see that  when
$\Omega_1^{(1)} = \Omega_1^{(2)}$ (solid lines in  Fig.~\ref{Cancel}),
the   magnetoconductivity   practically    does   not   change    when
$H^\prime_{SO}/H_\varphi$ changes from $0$ to $4$. This shows that the
two terms in linear splitting almost cancel each other in NMR, and the
result looks as if there were no linear splitting at all, even  though
the latter can be much larger  that the cubic splitting. On the  other
hand, the  same change  in $H^\prime_{SO}/H_\varphi$  has very  strong
effect on $\Delta \sigma$ when only one of the linear splitting  terms
is   present   {\bf(}$\Omega_1^{(2)}   =   0$,   dashed   lines{\bf)}.
Fig.~\ref{Cancel}b shows that the  same trend persists even  for large
cubic splittings,  though the  effect becomes  less dramatic.  We must
stress again that no such  cancellation occurs in the spin  relaxation
rates, which are sensitive only to the total spin splitting.

The cancellation discussed above  is more than an  abstract curiosity.
The Rashba term Eq.~(\ref{HamAsymm})  in quantum wells can  be changed
by  deformation.  For  a  $[001]$  quantum  well,  a deformation along
$(110)$ or $(1\bar{1}0)$ will, according to Eq.~(\ref{deform})  change
the  coefficient  $\alpha$  in  Eq.~(\ref{HamAsymm}).  The   resulting
splitting can exceed the contribution Eq.~(\ref{HamSymm}) for not  too
high deformations\cite{Pikus84}. Such  an experiment would  allow {\em
independent} measurement of the magnitudes of linear and cubic in  $k$
spin splittings Eq.~(\ref{Omega}), as  well as the sign  and magnitude
of the part  of the coefficient  $\alpha$, which is  determined by the
well asymmetry.

We should also note that the recent paper  Ref.~\onlinecite{Andrada94}
contains discussion of the contributions  of two types of linear  spin
splitting: the Rashba  term and the  Dresselhaus term. The  authors of
Ref.~\onlinecite{Andrada94}   have    used   spin-orbit    splittings,
calculated in  Ref.~\onlinecite{Dresselhaus92}. These  splittings were
derived  from   the  experimental   data,  using   the  formulas  from
Ref.~\onlinecite{Altshuler81},   which   implies   an  assumtion  that
splitting of both types give additive contributions to NMR, similar to
their  contribution  to  spin  relaxation  time. The results presented
above  show  that  in  fact  the  situation  is  exactly opposite: the
appearance  of  splitting  of  the  second type decreases, rather than
increases, the  total contribution  of linear  splitting to  NMR. This
contribution  continues  to  decrease  until  both  terms  in   linear
splitting becomes equal.

As far as  comparison of theory  and experiment is  concerned, no good
agreement had been achieved for quantum wells (unlike the NMR in metal
films,  where  the  theory  provides  very  accurate  description   of
experimental         results).          The         theory          of
Refs.~\onlinecite{Altshuler81,Lommer88}  was  unable  to  describe the
experimental results of Refs.~\onlinecite{Dresselhaus92,Hansen93} in a
wide range of magnetic fields\cite{Andrada94,Dresselhaus92}. The  main
reason  for  the  discrepancy  between  experiment  and theory was the
assumption that linear and cubic terms give additive contributions  to
the magnetoresistance and the formula of Ref.~\onlinecite{Altshuler81}
can be used  for the D'yakonov--Perel'  spin relaxation mechanism.  It
was first shown in Ref.~\onlinecite{Iordanskii94} that this assumption
is incorrect; however, no comparison with experiment was presented  in
this paper. We now proceed to  illustrate that the new theory is  able
to describe magnetoconductivity  in semiconductor quantum  wells quite
accurately.

\begin{figure}[t]

\epsfxsize=3 in
\epsffile{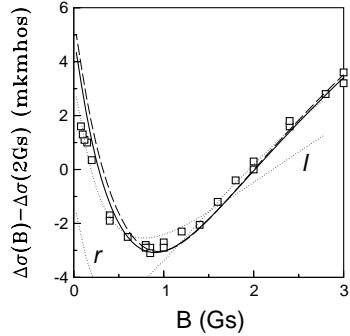}

\caption{Comparison of  theoretical and  experimental results  for the
magnetoconductivity. Squares show the experimental results of Ref.~18.
Solid  line  shows  the  best  fit  obtained  from our theory, fitting
parameters   are   $H_\varphi = 0.028\;{\rm Gs}$,   $H_{SO}/H_\varphi
=   31$,
$H^\prime_{SO}/H_\varphi=28.5$, and $H^{(1)}_{SO}/H_\varphi = 26$. The
best fit with only one term in linear splitting is shown by the dashed
line,   fitting   parameters   are   $H_{SO}/H_\varphi   =   26$   and
$H^\prime_{SO}/H_\varphi=23.5$. The fits for $\Omega_1 = 0$ are  shown
by doted lines. It is not possible to fit the experiment in the  whole
range of magnetic fields in this case. The fit which works best to the
left  of   the  magnetoconductivity   minimum  (marked   by  $l$)  has
$H_{SO}/H_\varphi  =  6.7$ (this is the value found in Ref.~18),  the
fit  to
the right side of the minimum
(marked by $r$) has $H_{SO}/H_\varphi = 4$. \label{Experiment}}

\end{figure}

In Fig.~\ref{Experiment}  we show  the comparison  of the experimental
results from Ref.~\onlinecite{Dresselhaus92} with the theory presented
in this  paper. The  main difficulty  in obtaining  a well-defined fit
arises from the cancellation of the two terms in the linear splitting,
discussed above. Indeed, the best fit, shown in  Fig.~\ref{Experiment}
by  the  solid  line,  is  obtained  for  ${\Omega_1^{(1)}}^2   \tau_1
\tau_\varphi = 13$,  ${\Omega_1^{(2)}}^2 \tau_1 \tau_\varphi  = 1.25$,
and ${\Omega_3}^2 \tau_3  \tau_\varphi = 1.25$.  If one wants  to find
the  best  fit  with  only  one  term  in  the linear splitting of the
conduction band, the fitting parameters are ${\Omega_1^{(1)}}^2 \tau_1
\tau_\varphi =  11.75$, $\Omega_1^{(2)}=0$,  and ${\Omega_3}^2  \tau_3
\tau_\varphi = 1.25$, and the agreement is also very good.  Comparison
of the two sets of fitting parameters above shows that the addition of
the second  term in  linear splitting,  with $\Omega_1^{(2)}$,  almost
cancels a part of the first term, with $\Omega_1^{(1)}$. The effect is
that  the  main  dependence  of  the magnetoconductivity on the linear
splitting can be described by the parameter  $\left({\Omega_1^{(1)}}^2
-  {\Omega_1^{(2)}}^2\right)  \tau_1   \tau_\varphi$,  and  an   equal
increase of  both $\Omega_1^{(1)}$  and $\Omega_1^{(2)}$  makes only a
small difference. On the other hand, an attempt to fit the  experiment
with the formula of  Ref.~\onlinecite{Altshuler81}, which can be  used
for $\Omega_1=0$, fails: one can see in Fig.~\ref{Experiment} that  it
is possible to fit the magnetoconductivity either on the right of  the
minimum or on  the left, but  not in the  whole range of  the magnetic
field.  The  cancellation  of  the  linear  splitting,  shown   above,
emphasizes the importance of magnetoconductivity measurements under  a
deformation, where  the ratio  of the  linear splitting  terms can  be
controlled independently.

Using values  of the  characteristic magnetic  fields Eq.~(\ref{HSO}),
determined from the fit, we can estimate the coefficients $\gamma$ and
$\alpha$         of         the         spin-orbit         Hamiltonian
Eqs.~(\ref{HamBulk},\ref{HamAsymm}). From Eq.~(\ref{Omega}),

\begin{equation}
\gamma = {\hbar \Omega_1^{(1)} \over k \left( \left\langle k_z^2
\right\rangle - {1 \over 4} k^2 \right)}.
\label{gamma}
\end{equation}

\noindent Here we should take $k$ equal to the Fermi wave vector  $k_F
= \sqrt{\,2 \pi N_s}$,  and $\left\langle k_z^2 \right\rangle$  can be
estimated using  the Fang-Howard  wave function  \cite{Fang66} for the
electrons in the heterostructure:

\begin{equation}
\psi(z) = \sqrt{\,{b^3 \over 2}}\,z\: {\rm e}^{-bz/2}.
\label{psi}
\end{equation}

\noindent  Then  $\left\langle  k_z^2  \right\rangle  =  b^2/4$.   The
parameter $b$ is determined mainly by the density of the electron gas.
We  can  estimate   $b$,  using  the   simple  expression,  given   in
Ref.~\onlinecite{Ando82}:

\begin{equation}
b = \left({16.5 \pi e^2 m N_s \over \kappa \hbar^2} \right)^{1/3}\!\!\!\!,
\end{equation}

\noindent where  $\kappa$ is  the dielectric  constant and  $m$ is the
electron effective mass.

From   the   fit   in   Fig.~\ref{Experiment}   we   have   the  value
${\Omega_1^{(1)}}^2  \tau_1  \tau_\varphi  \approx  13$.  The  product
$\tau_1 \tau_\varphi$  can be  found from  the value  of the  magnetic
field  $H_\varphi  \approx  0.028  \;{\rm  Gs}$, because, according to
Eq.~(\ref{HSO}),

\begin{equation}
H_\varphi = {c \hbar \over 4 e D \tau_\varphi} =
{c \hbar \over 2 v_F^2 e \tau_1 \tau_\varphi},
\end{equation}

\noindent  where  $v_F  =  \hbar  k_F/m$  is  the  Fermi  velocity  of
electrons.  Using  the  electron  density  $N_s  =  6.1\:10^{11}\;{\rm
cm}^{-2}$   from   Table~1   of  Ref.~\onlinecite{Dresselhaus92},  the
electron mass in GaAs $m = 0.067 m_0$, and the dielectric constant  of
GaAs $\kappa = 12.55$, we obtain the following estimate:

\begin{equation}
\gamma \approx 25\; {\rm eV \;\AA^3}.
\end{equation}

\noindent This value  for $\gamma$ agrees  surprisingly well with  the
results, obtained in  Ref.~\onlinecite{Pikus84} from the  measurements
of spin relaxation using optical orientation.

The fit also allows to estimate the coefficient $\alpha$ of the Rashba
term:

\begin{equation}
\alpha = {\hbar \Omega_1^{(2)} \over k_F}.
\end{equation}

\noindent  Using  the  value  ${\Omega_1^{(1)}}^2  \tau_1 \tau_\varphi
\approx 1.25$ we get the estimate

\begin{equation}
\alpha \approx 1.2\; {\rm meV \;\AA}.
\end{equation}

\noindent  The  value  of  the  coefficient  $\alpha$  had  never been
measured, and, as  we noted before,  it would have  been exactly 0  in
non-deformed quantum  wells if  the effective  mass approximation were
working   everywhere,    including   the    interface.   Authors    of
Refs.~\onlinecite{Lommer88,Andrada94} have estimated this coefficient,
assuming  that  the  interface  gives  no  contribution  at all, for a
uniform electric field in the quantum well:

\begin{equation}
\alpha_{\rm theor} = {\hbar^2 \over 2 m} {\Delta \over E_g}
{2 E_g + \Delta \over (E_g + \Delta) (3 E_g + 2 \Delta)} \,
e E,
\label{alphacalc}
\end{equation}

\noindent  where  $E_g$  is  the  direct  band  gap,  $\Delta$  is the
spin-orbit  energy  splitting,  $E$  is  the  electric  field.  In   a
heterostructure the electric field  changes in the $z$  direction from
$4 \pi N_s e/\kappa$  at the interface to  practically 0 on the  other
side of the  electron gas. In  this case it  should be replaced  by an
average electric field

\begin{equation}
\langle E \rangle = \int\limits_0^\infty E(z)\, |\psi(z)|^2 \ dz,
\label{avE}
\end{equation}

\noindent where  $E(z)$ itself  is determined  by the  distribution of
electron density:

\begin{equation}
E(z) = {4 \pi e N_s \over \kappa} \left(1 - \int\limits_0^z
|\psi(z^\prime)|^2 \ dz^\prime \right).
\label{Ez}
\end{equation}

\noindent We have  taken the following  values for the  energy gaps of
GaAs:\cite{RefBook} $E_g = 1.42\; {\rm eV}$ and $\Delta = 0.33\;  {\rm
eV}$.     Substitution      of     Eqs.~(\ref{avE},\ref{Ez})      into
Eq.~(\ref{alphacalc}) yields the estimate

\begin{equation}
\alpha_{\rm theor} = 2.2\; {\rm meV \;\AA}.
\end{equation}

\noindent  This  number  is  about  twice  the  value  which  fits the
experiment. We believe this is quite reasonable agreement  considering
the fact that  the estimate Eq.~(\ref{alphacalc})  is really an  upper
bound,  because  it  neglects  the  contribution  of  the field in the
interface, and  the latter  tends to  decrease $\alpha$.  On the other
hand, if one  assumes $\alpha =  2.2\; {\rm meV  \;\AA}$ and uses  the
cancellation  effect  discussed  above  to  add  equal  corrections to
$\Omega_1^{(1)}$  and  $\Omega_1^{(2)}$,  this  will  result  in a fit
nearly as  good as  the one  we suggested.  For this  new fit $\gamma$
changes to approximately  $28\; {\rm eV  \;\AA^3}$, and the  change is
well within the accuracy of existing determination of $\gamma$.

Finally, we  can use  the value  of $\Omega_3^2  \tau_3 \tau_\varphi =
1.25$ to estimate the ratio $\tau_3/\tau_1$. From Eq.~(\ref{Omega}) it
follows that

\begin{equation}
{\Omega_3 \over \Omega_1^{(1)}} = {4 \left( \left\langle k_z^2
\right\rangle -
{1 \over 4} k_F^2 \right) \over k_F^2},
\end{equation}

\noindent which gives the value

\begin{equation}
{\tau_3 \over \tau_1} \approx {1 \over 4}.
\end{equation}

\noindent  Theoretically,  this  ratio  can  vary  from  $1/9$\/   for
small-angle scattering (like scattering on remote charged  impurities)
to $1$\/ for short-range scattering. For a structure with fairly large
mobility, $\mu  \approx 1.1  \:10^5\; {\rm  cm}^2/{\rm V  \;sec}$, the
value $1/4$\/ is not unreasonable.

In conclusion, we  have presented a  new, improved theory  for quantum
interference corrections to the conductivity  of an electron gas in  a
semiconductor quantum well  in a magnetic  field. The theory  is valid
for D'yakonov--Perel'  spin relaxation  and when  the phase relaxation
time $\tau_\varphi$ is much  longer than the momentum  relaxation time
$\tau_1$, so that the diffusion approximation can be used. Our  theory
correctly takes into account  the contributions of different  terms in
spin splitting of  the conduction band.  We have shown  that while the
spin relaxation rate depends only  on the total magnitude of  the spin
splitting, the different parts  of the latter give  {\em non-additive}
contributions into the  magnetoresistance. Furthermore, the  two terms
in the  linear in  wave vector  part of  the spin  splitting, known as
Rashba and Dresselhaus  terms, actually cancel  each other when  their
magnitudes are comparable.  Using the new  theory we were  able to fit
the experimental data for the  magnetoconductivity in a wide range  of
magnetic  fields.  The  spin--orbit  splitting  coefficient  for   the
conduction band, obtained from the fit, is in very good agreement with
the one measured in optical orientation experiments.

Lastly, we would like to note that the spin splitting leads to similar
interference   corrections    to   magnetoconductivity    of   hopping
conductors\cite{Raikh94}.  We  expect  that  in  this  case it is also
important  to  distinguish  between   different  terms  in  the   spin
Hamiltonian, whose contributions to NMR will not be additive.

We are  grateful to  J. Allen  and D.  Hone for  helpful comments. The
authors acknowledge  support by  the San  Diego Supercomputer  Center,
where  part  of  the  calculations  were  performed.  The research was
supported in part by the Soros Foundation (G. E. P.) and by the Center
for Quantized Electronic Structures (QUEST) of UCSB (F. G. P.).



\end{document}